\documentclass[twocolumn,prl]{revtex4}
\usepackage{graphicx}    
\usepackage{dcolumn}     
\usepackage{bm}          
\usepackage{amsmath}
\usepackage{amssymb}

\begin{document}

\title{Theory of tunneling spectroscopy in the Larkin-Ovchinnikov state}

\author{Y.~Tanaka$^{1,2}$, Y.~Asano$^{3}$, M.~Ichioka$^{4}$ and S.~Kashiwaya$^{5}$ }
%
\affiliation{
$^1$ Department of Applied Physics, Nagoya University, Nagoya, 464-8603, Japan \\
$^2$ CREST, Japan Science and Technology Corporation (JST) Nagoya, 464-8603, Japan \\
$^3$ Department of Applied Physics, Hokkaido University, Sapporo 060-8628, Japan\\
$^4$ Department of Physics, Okayama University, Okayama 700-8530, Japan \\
$^5$ NeRI of National Institute of Advanced Industrial Science and
Technology (AIST), Tsukuba, 305-8568, Japan \\
}

\begin{abstract}
A theory of the tunneling spectroscopy of normal metal / 
Larkin-Ovchinnikov (LO) state is presented by fully taking account of the 
periodic modulation of pair potentials in real space. 
The resulting tunneling spectra have characteristic 
line shapes with several maxima and minima reflecting minigap 
structures due to the periodic pair potentials. 
These features are qualitatively different from 
tunneling spectra of Fulde-Ferrell state and those of uniform $s$- and $d$-wave 
superconductors.
On the basis of calculated results, we propose a method to
identify the LO state in actual experiments. 
\end{abstract}
\maketitle
Fulde and Ferrell~\cite{Fulde} and Larkin and
Ovchinnikov~\cite{Larkin} proposed superconducting 
states in which spin-singlet superconducting pair potentials are  
periodically modulated in real space under
high magnetic fields.
%
%
%
The Zeeman spin splitting results in 
 a total momentum $2\boldsymbol{q}$ of Cooper pairs.
Fulde-Ferrell (FF) discussed that pair potential becomes
$\Delta\exp(i\boldsymbol{q\cdot r})$.
%
Larkin and Ovchinnikov (LO) proposed
independently an alternative scenario in which the order parameter
is real, but varies periodically in real space like
$\Delta\cos(\boldsymbol{q\cdot r})$~\cite{Larkin}. 
The two states are collectively called 
Fulde-Ferrell-Larkin-Ovchinnikov (FFLO) state.
Only a few attempts have been made at FFLO state~\cite{tachiki}
in the last century.
%
%
However FFLO state suddenly becomes a hot topic because evidence of FFLO 
state has been reported in a number of experiments
~\cite{radovan,bianchi2003,watanabe,kakuyanagi}
and theories~\cite{Houzet,Adachi,mizushima}.
These studies suggest realization of FFLO state in a heavy fermionic compound 
CeCoIn$_5$. 
According to the phase diagram of CeCoIn$_5$, LO state is considered to be 
more stable than FF state. 
To understand basic properties of LO state,
it is important to study detailed structures of quasiparticle energy spectra. 
In fact, experimental research in this direction becomes accessible 
now~\cite{Wei,Park}. 
%
Thus a theory of the tunneling spectroscopy of LO state is desired to 
interpret the tunnel spectra in experiments.

In the presence of the spatial modulation of 
pair potentials~\cite{Tsukada91,Vanevic}, 
local density of states (LDOS) is 
expected to be very different from that in the conventional $U$-shaped subgap 
structure in uniform superconductors.
Up to now, however, there is no theory of tunneling spectroscopy in which effects of
the periodic sign change of pair potential in real space on spectra are fully taken into account.
In addition, we also have to consider effects of sign change of pair potential in $\boldsymbol{k}$-space 
because CeCoIn$_{5}$ is a strongly correlated material and its
promising pair potential is $d$-wave symmetry. 
In such a pairing symmetry, charge transport of the 
junctions are governed by a mid gap Andreev resonant state (MARS) formed at 
the junction interface~\cite{hu1994,tanaka,kashiwaya,lofwander,wei1998}.
Thus we must take into account the two kinds of sign change in pair potentials, 
i.e., the sign change in real space and that in $\boldsymbol{k}$-space, on an equal footing. 

In this  paper, we present a theory of 
tunneling spectra in normal metal / insulator /
superconductor in LO state (N/I/LO) for the first time. 
We choose that $\boldsymbol{q}$ is parallel to the normal of the
junction interface. 
In such junctions, tunneling spectra are expected to have fine structures 
reflecting the miniband due to the spatial modulation of pair potentials.  
We consider two types of junction: 
(1) \textit{node contact junctions} where 
the amplitude of LO pair potentials vanishes at the junction 
interface and (2) \textit{maximum contact junctions} where the amplitude of 
pair potentials takes its maximum at the junction interface. 
In addition to tunneling spectra, we calculate LDOS of bulk LO state
at two local points: one is a place at which pair potential takes 
its maximum (maximum point) 
and the other is a place at which pair potential has a node (nodal point). 
Calculated results show complex structures in tunneling spectra and LDOS 
depending on types of junction and pairing symmetries.
These are a consequence of characteristic quasiparticle excitation in LO state such as
miniband structures, Zeeman spin splitting and MARS at the interface.
By comparing tunneling spectra and LDOS, we discuss a relation between
quasiparticle states originating from the sign change of pair potential in 
$\boldsymbol{k}$-space and those originating from the sign change in real space.
%
%
%
%
The calculated results give us useful information to
identify the pairing symmetry of LO state in CeCoIn$_{5}$. \par

Let us consider an N/I/LO junctions in two-dimension
as shown in Fig.~1(a), where N and LO correspond to regions for $x<0$ and $x>0$, respectively. 
We assume a flat interface in the $y$ direction and isotropic Fermi surface.
Magnetic field $H$ is applied in the $x$ direction so that effects of magnetic fields on
orbital part of wave function can be neglected.
The insulating barrier is expressed by the 
delta-function model and its potential is given by $H_{b}\delta(x)$.  
The effective mass $m$ and Fermi wave number $k_{F}$ are chosen to be common in N and LO. 
Here, we neglect the spin-orbit coupling in LO. 
The tunneling conductance in the junctions $\sigma_{S}$ 
at zero temperature can be expressed by 
the summation of up and down components with 
\begin{align}
\sigma_{S}(eV)=&(\sigma_{\uparrow} + \sigma_{\downarrow})/2,\\
\sigma_{\uparrow(\downarrow)}
=&\int^{\pi/2}_{-\pi/2} d\theta \; \cos\theta\; \sigma_{N} \; F_{\uparrow(\downarrow)}, \\
F_{\uparrow(\downarrow)}
=&\frac{1 + \sigma_{N}\mid \Gamma_{+\uparrow(\downarrow)}\mid^{2}
+ (\sigma_{N}-1)\mid \Gamma_{+\uparrow(\downarrow)} 
\Gamma_{-\uparrow(\downarrow)} \mid^{2} }
{\mid 1 
+ (\sigma_{N} -1)\Gamma_{+\uparrow(\downarrow)}\Gamma_{-\uparrow(\downarrow)} 
\mid^{2}},\label{defgamma}
\end{align}
with $V$ being a bias voltage~\cite{kashiwaya}.
The transparency at the interface is given by $\sigma_{N}(\theta)=
4\cos^{2}\theta/(4\cos^{2}\theta + Z^{2})$,  
where $Z=2mH_{b}/\hbar k_{Fx}$, 
$\theta$ is an injection angle of a quasiparticle at the 
interface measured from the $x$ direction and $k_{Fx}$ is the $x$ component of 
the Fermi wave vector. 
%
In the above, $\Gamma_{\pm\uparrow}$ and $\Gamma_{\pm\downarrow}$ 
obey the following equations
\begin{align}
-iv_{Fx}\partial_{x} \Gamma_{\pm\uparrow}
=&\Delta(\theta_{\pm},x)(1+\Gamma_{\pm\uparrow}^{2})-2\Gamma_{\pm\uparrow}
(\varepsilon + \frac{\omega_{L}}{2}),\\
-iv_{Fx}\partial_{x} \Gamma_{\pm\downarrow}
=&\Delta(\theta_{\pm},x)(1+\Gamma_{\pm\downarrow}^{2})-2\Gamma_{\pm\downarrow}
(\varepsilon - \frac{\omega_{L}}{2}),
\end{align}
where $\theta_{+}=\theta$, $\theta_{-}=\pi-\theta$, 
$\varepsilon$ denotes a quasiparticle energy measured from the 
Fermi energy, 
$\omega_{L}$ is the spin Larmor frequency, and 
$v_{Fx}$ is the $x$ component of the Fermi velocity, respectively. 
The spatial dependence of pair potential 
$\Delta(\theta,x)$ is given by 
$\Delta(\theta,x)=\Delta(x)\Theta(x)f(\theta)$, 
where $\Theta(x)$ is a step function. 
The form factor $f(\theta)$ depends on pairing symmetries
and orientations of crystalline axis in CeCoIn$_5$: 
$f(\theta)=1$ for $s$-wave symmetry, 
$f(\theta)=\cos 2\theta$ for $d_{x^2-y^2}$-wave symmetry, 
and $f(\theta)=\sin 2\theta$ for $d_{xy}$-wave symmetry. 
We calculate $\Gamma_{\pm\uparrow(\downarrow)}$ in Eq.~(\ref{defgamma}) from
$2 \times 2$ quasiclassical Green's functions 
based on the evolution operator method~\cite{Matsuo}. 
%
We choose the value $\omega_{L}$ as 
$\omega_{L}=0.8\pi T_{C}$ with $T_{C}$ being the transition temperature. 
The resulting effective magnitude of the 
Zeeman splitting is about $0.7\Delta_{0}$ with $\Delta_{0} \sim 0.56\pi T_{C}$.
%
%
In the present paper, we describe the spatial dependence of LO pair potentials 
by $\Delta(x)=\Delta_{0}\cos(Q x + Q_{1})$, where $Q=2\pi /L$ with $L$ being 
the period of oscillations in pair potentials. 
The period $L$ is measured in units of $L_{0}=2\pi/Q_{0}$ 
with $Q_{0}=\frac{\pi T_{C}}{\hbar v_{F}}$. 
The barrier parameter $Z$ at the interface 
is chosen to be $Z=5$ throughout this paper, and resulting transparency of junction is about 0.1. 
In what follows, we discuss normalized tunneling conductance
$\sigma_{T} = \sigma_{S}(eV)/\bar{\sigma}_N$, where $\bar{\sigma}_N=\int d\theta\; \cos\theta
\sigma_N $ is the normal conductance
of junctions. In addition to $\sigma_T$, we also calculate $\rho_{T}$ which is LDOS normalized 
 by its value in the normal state. In experiments, LDOS can be measured in scanning
tunneling spectroscopy as shown in Fig.~1(b).
\par

%

\begin{figure}
\begin{center}
\includegraphics[width =7.5cm]{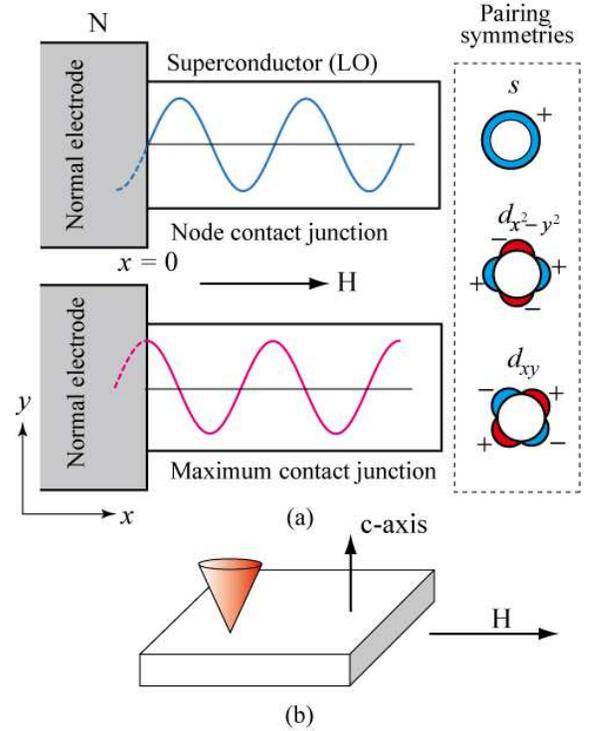}
\end{center}
\caption{\label{fig:fig1} 
A normal metal / insulator / superconductor in LO state junction is shown in (a).
In a node contact junction and a maximum contact junction,  
$Q_{1}$ is set to be $-\pi/2$ and 0, respectively. 
A schematic illustration of scanning tunneling spectroscopy is in (b),
where a cone represents a scanning probe for local density of states. }
\end{figure}
First we focus on the $s$-wave symmetry. 
In Fig.~2(a) and (c), $\sigma_{T}$ of the node contact junction and that of the 
maximum contact junction are shown, where we choose period of 
oscillations to be $L=L_0$ for dotted line $a$ 
and $L=10L_0$ for solid line $b$. 
In Figs.~2(b) and (d), $\rho_T$ at a nodal point and that at a maximum point of
bulk superconductors are shown.
The LDOS at the nodal and maximum points of the pair potential 
are significantly different from each other~\cite{Vorontsov}
as shown in Figs.~2(b) and (d). 
In both types of junctions, $\sigma_{T}$ and $\rho_T$ have 
fine structures reflecting the miniband structures 
due to spatial oscillations of pair potentials. 
When we compare curve $a$ with curve $b$,
fine structures are more remarkable in longer period of oscillations.
In the maximum contact junctions, line 
shapes of $\sigma_{T}$ and those of $\rho_T$ are very similar to each other. 
This feature becomes clearer for longer period of oscillations. 
On the other hand in the node contact junctions, 
line shapes of $\sigma_{T}$ are very different from 
those of $\rho_T$. 
Here we note that results for $d_{x^2-y^2}$-wave symmetry are 
qualitatively the same as those in the $s$-wave junctions in Fig.~2. 

\begin{figure}
\begin{center}
\includegraphics[width=8cm]{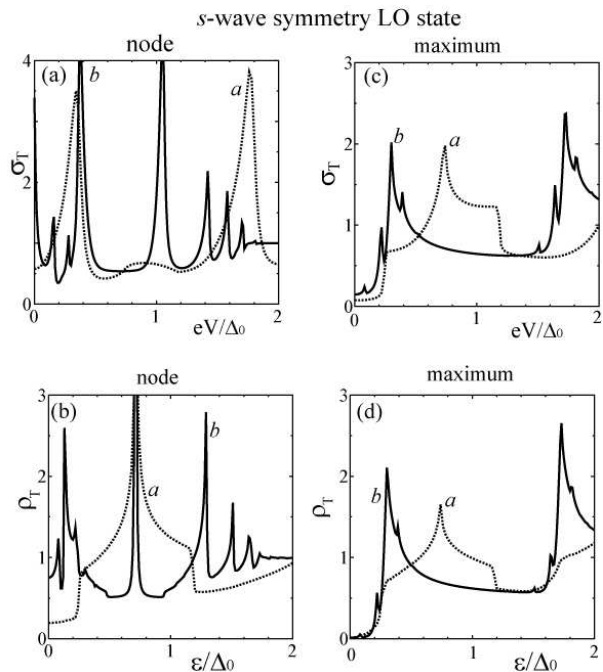}
\end{center}
\caption{ Calculated results for $s$-wave N/I/LO junctions.
Normalized tunneling conductance $\sigma_{T}$ is plotted as a function of bias voltages 
for the node contact junction in (a) and for the maximum contact junction in (c).
Local density of states $\rho_T$ at a nodal point and that at a maximum
point of bulk LO state are shown in (b) and (c), respectively. 
In all panels, $a$: $L=L_{0}$ and $b$: $L=10L_{0}$.
}
\label{fig:2}
\end{figure}
Secondly we focus on $d_{xy}$-wave N/I/LO junctions.
In this case, MARS is formed at the interface due to the 
sign change of the pair potentials in $\boldsymbol{k}$-space.
When spatial modulation of pair potentials is absent, 
$\sigma_{T}$ has a splitting peaks around the zero bias. 
The splitting width corresponds to the Zeeman splitting energy.
We show calculated results in Figs.~3(a)-(d).
The line shapes of the $\sigma_{T}$ in the maximum contact junction in (c)
are very much different from those of $\rho_{T}$ in (d).
On the other hand in the node contact junctions, the line shapes of the 
$\sigma_{T}$ in (a) are similar to those of $\rho_{T}$ in (b). 
The correspondence between $\sigma_{T}$ and $\rho_{T}$ 
is tending to be clearer for the longer period LO state.
A number of peaks and dips due to the miniband structures can be seen in curve 
$b$ in Figs.~3(a) and (b).
\par

\begin{figure}
\begin{center}
\includegraphics[width=8cm]{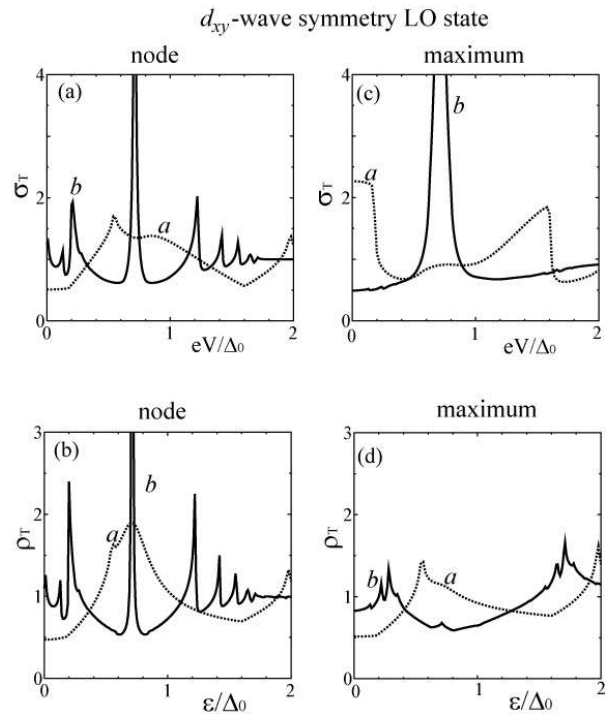}
\end{center}
\caption{\label{fig:pfig3} 
Calculated results for $d_{xy}$-wave N/I/LO junctions.
}
\end{figure}

We summarize obtained results in Figs.~2 and 3, and explain 
the characteristic features by using schematic pictures in Fig.~4.  
The sign change of pair potential in $\boldsymbol{k}$-space affects drastically
$\sigma_T$ and that in real space modifies $\rho_T$.
In the maximum contact junctions of $s$-wave symmetry, 
$\sigma_{T}$ well corresponds to $\rho_T$ because
the sign change of pair potential at the 
interface is absent in both $\boldsymbol{k}$-space and real space.
On the other hand in the node contact junctions of $d_{xy}$-symmetry, 
$\sigma_{T}$ well corresponds to $\rho_T$.
This is because $\sigma_T$ and $\rho_T$ reflect the sign change in $\boldsymbol{k}$-space and
that in real space, respectively. Thus tunneling spectra are closely related to 
LDOS even in the presence of MARS at the interface.
These features are specific to LO state. 
\par
\begin{figure}
\begin{center}
\includegraphics[width=8cm]{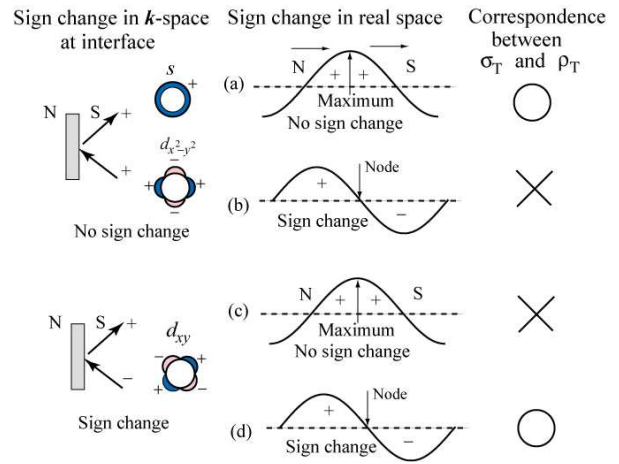}
\end{center}
\caption{\label{fig:pfig4} 
Schematic illustration of the sign change of the pair potential.
Correspondence between the  
reflection process at the interface and the 
local density of states in the bulk LO state. 
}
\end{figure}

Thirdly we look into $\sigma_{T}$ for FF state as shown in Fig.~5. 
In FF state, pair amplitudes are uniform in real space.
Thus a preexistence formula for uniform superconductors~\cite{tanaka}
can be applied to the N/I/FF junctions even in a presence of the Zeeman splitting.
The calculated results of $\rho_{T}$ do not have a spatial dependence 
contrary to those of LO state. 
For $s$-wave case, the line shape of $\sigma_{T}$ 
becomes similar to that of $\rho_{T}$ for longer period FF state 
(see curve $b$ in (a) and (b)). 
On the other hand, for $d_{xy}$-wave case, 
the line shape of $\sigma_{T}$ 
has a splitting zero-bias conductance peak (ZBCP) 
due to the sign change of pair potentials in $\boldsymbol{k}$-space~\cite{Cui}. 
For almost all of the cases, 
the overall features are very different from those in LO junctions. 
Only for the case of FF state with $d_{xy}$ pairing, 
the line shape of $\sigma_{T}$ of longer period junctions is similar to 
the corresponding case of LO state. 
%
%
\begin{figure}
\begin{center}
\includegraphics[width=8cm]{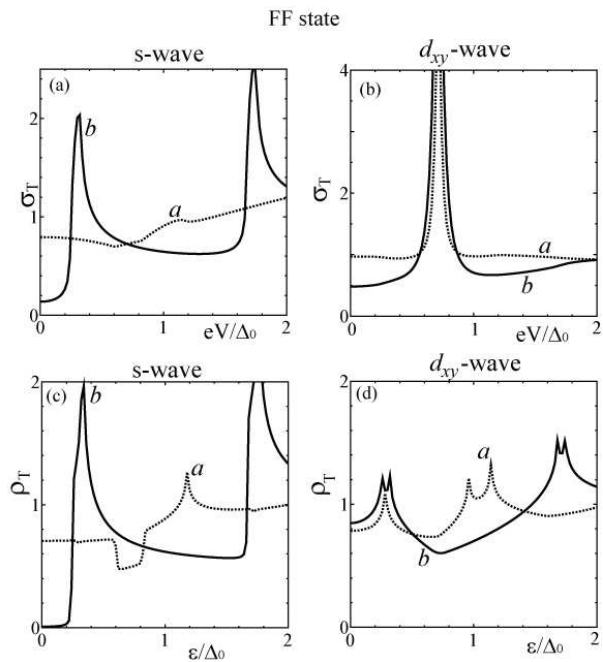}
\end{center}
\caption{\label{fig:pfig5} 
Calculated results for N/I/FF junctions.
Normalized tunneling conductance $\sigma_{T}$ is plotted as a function of bias voltages 
for $s$-wave in (a) and that for $d_{xy}$-wave in (c).
Local density of states $\rho_T$ for $s$-wave and that for $d_{xy}$-wave
are shown in (b) and (c), respectively. 
%
}
\end{figure}

Finally we propose a tunneling experiment at a sufficiently low temperature 
and predict observed tunnel conductance
on the basis of the obtained results. 
Since CeCoIn$_{5}$ is a strongly correlated electron system, 
realization of the $d$-wave state is more plausible than that of $s$-wave 
\cite{Izawa,Nishikawa}. 
If the (110) crystalline axis of CeCoIn$_5$ is perpendicular to junction interface, 
which corresponds to $d_{xy}$-wave symmetry in this paper,
a ZBCP is expected in tunneling spectra in the absence of magnetic field. 
We note that ZBCP can be always expected in $d$-wave junction except for 
an idealistic junction in which (100) crystalline axis of CeCoIn$_5$ is 
just perpendicular to junction interface. 
%
%
At the zero magnetic field, $\sigma_{T}$ has a ZBCP. 
The height of ZBCP would decrease with the increase of magnetic fields 
and the ZBCP would split into two peaks because of the Zeeman effect. 
The splitting width of the ZBCP increases with the increase of 
magnetic fields. 
When we increase magnetic fields further, 
CeCoIn$_{5}$ undergoes transition to the LO state. 
After the transition, fine structures with many peaks and dips 
are expected in $\sigma_{T}$ in addition to the splitting ZBCP. 
When CeCoIn$_{5}$ undergoes the transition to the normal state
at the critical magnetic field, 
the splitting ZBCP and the fine structures would disappear.  
If above change in tunneling spectra would be observed, 
it must be concrete evidence of the LO state. \par

In conclusion, a theory of the tunneling spectroscopy of normal metal / 
Larkin-Ovchinnikov (LO) state is presented by fully taking account of the 
periodic modulation of pair potentials in real space. 
The resulting normalized tunneling conductance  
has a line shape with several maxima and minima reflecting minigap 
structures of density of states. 
These features are not expected 
in the Fulde-Ferrell (FF) state because amplitude of pair potentials is
uniform in the FF state.
Our results serve as a guide to identify the LO state in the 
experiments. 


\end{document}